\documentclass[
aip,
pop,
amsmath, 
amssymb,
reprint,
floatfix,
]{revtex4-1}

\usepackage{graphicx}   
\usepackage{dcolumn}    
\usepackage{bm}         
\usepackage{siunitx}
\DeclareSIUnit{\angstrom}{\textup{\AA}} 
\usepackage{comment}
\usepackage[utf8]{inputenc}
\usepackage[T1]{fontenc}
\usepackage{mathptmx}
\usepackage{etoolbox}
\usepackage{float}
\usepackage{tabularray}
\usepackage[usenames,dvipsnames]{color}
\usepackage{hyperref}
\def\Snospace~{\S{}}

\usepackage[all]{hypcap}
\hypersetup{urlcolor=Blue, citecolor=Blue,linkcolor=Blue,colorlinks=true}

\newcommand{\sref}[2]{\hyperref[#1]{Fig. \ref{#1}#2}}

\begin{document}
\title[]{Radiative cooling effects on reverse shocks formed by magnetised supersonic plasma flows}
\author{S. Merlini}
\email[Email address: ]{s.merlini19@imperial.ac.uk}
\affiliation{Blackett Laboratory, Imperial College, London SW7 2AZ, United Kingdom}
\author{J. D. Hare}
\affiliation{Blackett Laboratory, Imperial College, London SW7 2AZ, United Kingdom}
\affiliation{Plasma Science and Fusion Center, Massachusetts Institute of Technology, Cambridge MA 02139, USA\looseness=-1}
\author{G. C. Burdiak}
\altaffiliation[Current address: ]{First Light Fusion Ltd., 10 Oxford Industrial Park, Yarnton, Kidlington OX5 1QU, United Kingdom}
\author{J. W. D. Halliday}
\altaffiliation[Current address: ]{Atomic and Laser Physics Group, University of Oxford, Oxford, OX1 3PU, United Kingdom}
\affiliation{Blackett Laboratory, Imperial College, London SW7 2AZ, United Kingdom}
\author{A. Ciardi}
\affiliation{Sorbonne Université, Observatoire de Paris, PSL Research University, LERMA, CNRS UMR 8112
75005 Paris, France\looseness=-1}
\author{J. P. Chittenden}
\affiliation{Blackett Laboratory, Imperial College, London SW7 2AZ, United Kingdom}
\author{T. Clayson}
\altaffiliation[Current address: ]{Magdrive Ltd., Unit 4 Bep-0 Building, Thomson Ave, Harwell Oxford, Didcot, OX11 0GD, United Kingdom}
\affiliation{Blackett Laboratory, Imperial College, London SW7 2AZ, United Kingdom}
\author{A. J. Crilly}
\affiliation{Blackett Laboratory, Imperial College, London SW7 2AZ, United Kingdom}
\author{S. J. Eardley}
\affiliation{Blackett Laboratory, Imperial College, London SW7 2AZ, United Kingdom}
\author{K. E. Marrow}
\affiliation{Blackett Laboratory, Imperial College, London SW7 2AZ, United Kingdom}
\author{D. R. Russell}
\altaffiliation[Current address: ]{Technische Universitaet Muenchen, Forschungs-Neutronenquelle Heinz Maier-Leibnitz,
Lichtenbergstrasse 1, D-85748 Garching, Germany}
\author{R. A. Smith}
\author{N. Stuart}
\author{L. G. Suttle}
\author{E. R. Tubman}
\affiliation{Blackett Laboratory, Imperial College, London SW7 2AZ, United Kingdom}
\author{V. Valenzuela-Villaseca}
\altaffiliation[Current address: ]{Department of Astrophysical Sciences, Princeton University, Princeton, NJ 08544, USA}
\author{T. W. O. Varnish}
\altaffiliation[Current address: ]{Plasma Science and Fusion Center, Massachusetts Institute of Technology, Cambridge MA 02139, USA}
\author{S. V. Lebedev}
\affiliation{Blackett Laboratory, Imperial College, London SW7 2AZ, United Kingdom}
\date{\today} 
\begin{abstract}
    We study the structure of reverse shocks formed by the collision of supersonic, magnetised plasma flows driven by an inverse (or exploding) wire array with a planar conducting obstacle. We observe that the structure of these reverse shocks varies dramatically with wire material, despite the similar upstream flow velocities and mass densities. For aluminium wire arrays, the shock is sharp and well-defined, consistent with magneto-hydrodynamic theory. In contrast, we do not observe a well-defined shock using tungsten wires, and instead we see a broad region dominated by density fluctuations on a wide range of spatial scales. We diagnose these two very different interactions using interferometry, Thomson scattering, shadowgraphy, and a newly developed imaging refractometer which is sensitive to small deflections of the probing laser corresponding to small-scale density perturbations. We conclude that the differences in shock structure are most likely due to radiative cooling instabilities which create small-scale density perturbations elongated along magnetic field lines in the tungsten plasma. These instabilities grow more slowly and are smoothed by thermal conduction in the aluminium plasma.
\end{abstract}

\maketitle
 
\section{\label{sec:Introduction} Introduction}

Shocks are well-studied and wide-spread phenomena in astrophysical and laboratory plasmas, in which the standard hydrodynamic results are further modified by the presence of magnetic fields, radiative cooling and transport, and kinetic effects, which can all affect the structure and stability of the shock. 
In recent years, the fundamental physical processes of magnetised shocks in plasmas have been studied in the laboratory using a range of experimental platforms, including magnetised collisionless shocks driven by lasers,\cite{Schaeffer2016} unmagnetised laser-driven plasma flows interacting with magnetized obstacles,\cite{Levesque2022a} supercritical perpendicular shocks in a theta-pinch configuration,\cite{Endrizzi2021} and in a wide range of pulsed-power-driven wire-array configurations.\cite{Lebedev2019,Suttle2019}

Inverse (or exploding) wire arrays \cite{Harvey-Thompson2009} driven by pulsed-power generators make good platforms for studying magnetised shock physics as they produce long-lasting, radially diverging, super-sonic and often super-Alfv\'enic magnetised plasma flows. The mass density and velocity of the plasma flows from wire arrays are only weakly dependent on the wire material,\cite{Chittenden2004} which means that changing the wire material (and hence the ion mass and plasma ionization state) can give very different plasma resistivity, ion-skin depth,\cite{Russell2021} ion-ion mean free path,\cite{Swadling2016} and radiative cooling rate.\cite{Hare2018a} This in turn makes it possible to study shocks in very different plasma regimes simply by changing the wire material, and in particular we can achieve significantly different radiative cooling functions by using either mid or high-Z elements for the wires, such as aluminium or tungsten.

Thermal instabilities with radiative cooling have long been considered important in astrophysical systems,\cite{Field1965, Hunter1970} and in astrophysical shocks, these instabilities may lead to oscillating shock fronts,\cite{Chevalier1982} relevant to highly magnetised, low-$\beta$ accretion flows onto Young Stellar Objects,\cite{Orlando2010, Matsakos2013} which are effectively one dimensional due to the confining magnetic field. However, analysis and simulations of these shock oscillations in two or three dimensions suggest that the shock will fragment into a large number of oscillating regions rather than oscillating as a quasi-one-dimensional system.\cite{Orlando2010, Reale2013, Matsakos2013}
Further work has looked into the role of the strength\cite{Drake2009} and the orientation of the magnetic field with respect to the accretion flow,\cite{Ramachandran2005} which can further change the stability of the system to shock oscillations.
These analytical results and numerical simulations make the role of radiative cooling instabilities in magnetised shocks an interesting direction for further investigation, and there has been some work on studying these cooling instabilities in pulsed-power driven plasma flows.\cite{Suzuki-Vidal2015}

    \begin{figure*}
            \centering    
            \includegraphics[width = 1\linewidth]{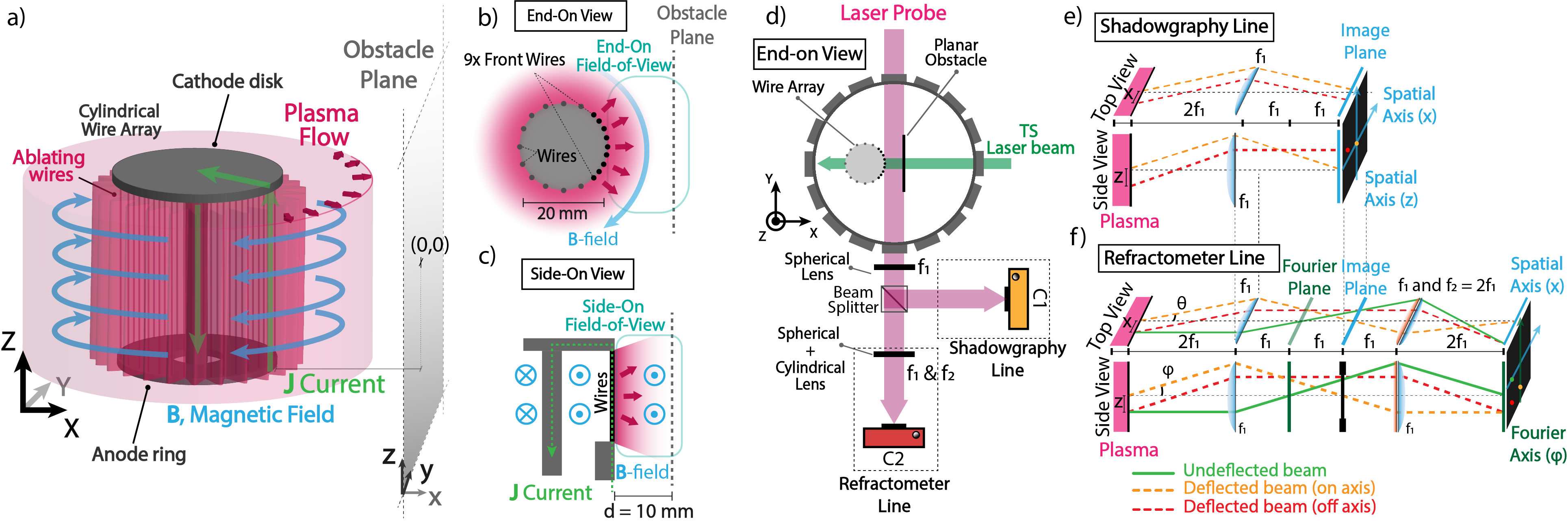}
            \caption{
            (a) Illustration of the exploding wire array experiment with planar obstacles. Obstacles are placed at a fixed distance from the plasma source, on the yz-plane perpendicular to the flow direction. (b) and (c) are the diagnostics projected views, respectively from the end-on and side-on. (d) Top-view schematic of the side-on diagnostic imaging system. After the first collecting lens, the laser probe is split between the shadowgraphy and refractometer lines using a beam splitter. (e) and (f) are ray diagrams of the shadowgraphy and refractometer imaging systems.
            }
            \label{fig:setup}
    \end{figure*}
        
This paper builds on previous work which used exploding aluminium wire arrays to study reverse shocks formed using planar obstacles,\cite{Lebedev2014} shocks formed around cylindrical obstacles aligned parallel and perpendicular to the advected magnetic field,\cite{Burdiak2017} shocks forming around magnetised obstacles,\cite{Suttle2019} sub-critical shocks around large, field-aligned obstacles,\cite{Russell2022a} and bow-shocks forming around small, pointed obstacles.\cite{Datta2022b} The principal novelty of this work is two-fold: we study reverse shocks formed with planar obstacles as in Ref. \onlinecite{Lebedev2014}, but we consider tungsten plasmas as well as aluminium plasmas, and we use a recently developed imaging refractometer diagnostic\cite{Hare2021} to make detailed measurements of the density perturbations upstream and downstream of the interaction region.

The paper is organised as follows: In Sec. \ref{sec:ExperimentalSetup} we discuss our experimental set-up, consisting of an exploding wire array and an obstacle, as well as the diagnostics used to measure the plasma parameters in these experiments -- interferometry, Thomson scattering, shadowgraphy and the imaging refractometer, and we include some experimental data to demonstrate the utility of the imaging refractometer.  In Sec. \ref{sec:ExperimentalResults} we present experimental results, starting with the conducting grid which produces a network of interacting shocks with a periodic structure, and then results from the collision of aluminium and tungsten plasma flows with a planar conducting wall, in which we observe very different behaviour depending on the wire material. We discuss these results in Sec. \ref{sec:Discussion}, and we consider several different mechanisms for the observed density perturbations. We conclude and present the outlook for future work in Sec. \ref{sec:Conclusion}.

\section{\label{sec:ExperimentalSetup}Experimental Setup and Diagnostics}

\subsection{An exploding wire array with an obstacle}\label{ssec:exploding_arrays}

The experiments presented in this paper were carried out using MAGPIE, a versatile 1.4 MA, 240 ns rise time, high-impedance pulsed-power generator.\cite{Mitchell1996} Previous studies have successfully demonstrated the capability of pulsed-power-driven inverse (or exploding) wire arrays to produce long-lasting plasma flows ($\approx 500$ ns) which advect a fraction of the driving magnetic field.\cite{Suttle2019} The experimental platform used in this study is illustrated in Fig. \ref{fig:setup}. The exploding wire array consisted of 20 thin metallic wires connected in parallel between two electrodes, forming a cylindrical arrangement around a central cathode. The array was 20 mm in diameter and 22 mm in height, and an obstacle was placed 10 mm away from the wires. In order to reduce the divergence of the flow in the region of interest, the 9 wires facing the obstacle were spaced by 11.25$^\circ$, with other wires spaced by 22.5$^\circ$.\cite{Burdiak2017}.
        
The absence of current return structures surrounding the array makes this platform easily accessible from both side-on and end-on (or axial) views for a suite of spatially and temporally resolved diagnostics, such as laser-probing interferometry,\cite{Swadling2014} optical Thomson scattering (TS),\cite{Suttle2021} and the new imaging refractometer,\cite{Hare2021} as shown in Fig. \ref{fig:setup}.  
  
\subsection{Diagnostics}\label{ssec:interferometry_and_ts}

Measurements of line-integrated electron density were made using a Mach-Zehnder imaging interferometer, similar to Ref. \onlinecite{Swadling2014}. Interferograms of the xy-plane were taken by probing along the end-on direction [Fig. \ref{fig:setup}(b)] using a 532 nm laser (Nd:YAG, 100 mJ, 500 ps), and were processed following the technique described in Refs. \onlinecite{Swadling2013} and \onlinecite{Hare2019}.

A spatially-resolved optical Thomson scattering (TS) diagnostic\cite{Suttle2021} was used to measure the flow velocity and to place an upper bound on the plasma temperature. A laser beam ($\lambda =$ 532 nm, 5 ns FWHM, 3 J) was focused through the plasma, and the scattered light was collected from 14 equally spaced positions using a lens and a linear array of optical fibers, with $150$ $\mu$m diameter collection volumes spaced by 0.3 mm.

\begin{figure*}
    \centering
    \includegraphics[width = 1\linewidth]{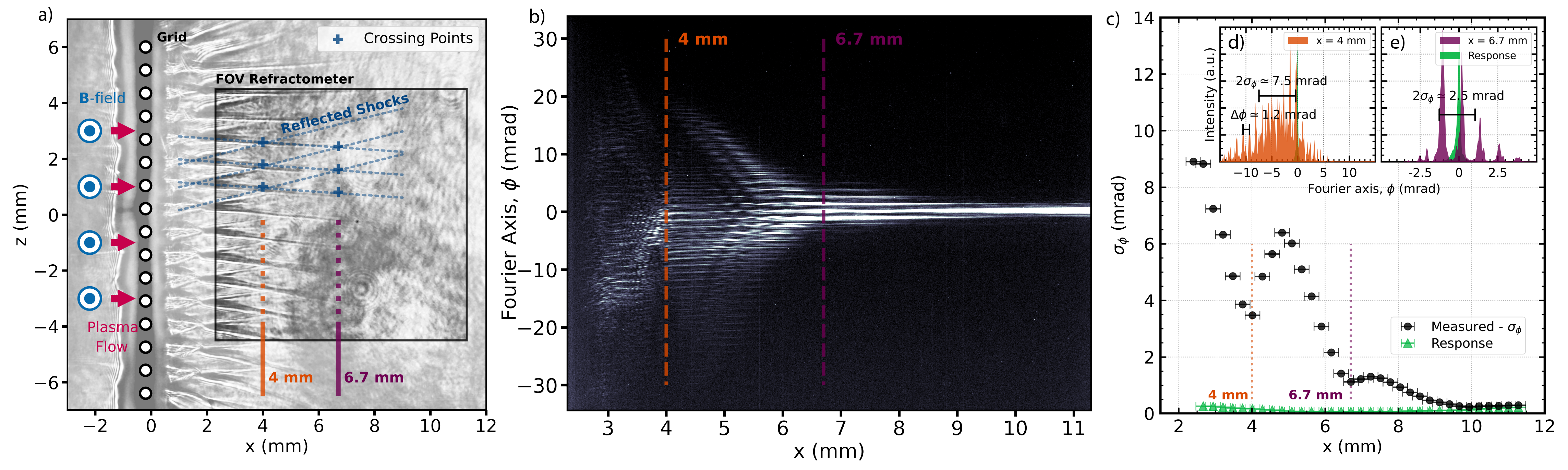}
    \caption{
    (a) 1053 nm side-on laser shadowgraphy image of the post-grid plasma flows at 404 ns after current start. The location of the origin is set at the position of the 1D grid and halfway up the wire array. Multiple interacting shocks are visible to the right of the grid.
    (b) Refractometer image of the highlighted Field-of-View (FOV) region in the shadowgraphy image, showing a broad envelope which decreases in angular width with increasing distance from the grid, with intense modulations within the envelope. In this figure, the image intensity has been adjusted to show the fine structure, but the raw image was not saturated.
    (c) The measured standard deviation of the refractometry image signal, $\sigma_\phi$, at each discrete x position, with $\Delta x = 0.4$ mm from a single experiment.
    (d), (e) Examples of angular intensity distribution respectively at x = 4 mm and x = 6.7 mm. $\Delta \phi \sim 1.2$ mrad.
    }

    \label{fig:grid_Al_shadow}
\end{figure*}

We also used a new diagnostic, the imaging refractometer, which is described in detail in Ref. \onlinecite{Hare2021}. The imaging system is shown in Fig. \ref{fig:setup}(d): the refractometer uses physical optics to simultaneously capture a conventional shadowgraphy image of the plasma and a hybrid image, which records the spatial position of the rays on the horizontal axis and the line-integrated deflection angle distribution of the rays on the vertical axis.
To demonstrate the difference between the shadowgraphy and the refractometer, we can consider a distribution function for the rays for a beam propagating in the y direction, $f(x,y,z, \theta, \phi)$, with $\theta$ and $\phi$ as the angle between y and x or z, respectively.
The laser probing beam incident on the plasma is collimated, so $f(x_0, y = 0, z_0, \theta = 0, \phi = 0)$.
After the plasma at $y = L_y$, the rays have picked up some deflection angle and have also moved in the xz plane due to these deflections, resulting in $f(x, y = L_y, z, \theta, \phi)$.
The shadowgraphy diagnostic measures the intensity distribution integrated over the two angular coordinates, $I(x, z, y = L_y) = \int\int f(x,y = L_y,z, \theta, \phi) d\theta d\phi$, whereas the imaging refractometer integrates over one spatial and one angular component and measures $I(x, \phi, y = L_y) = \int\int f(x,y = L_y, z, \theta, \phi) d\theta dz$.
As such, the output of the imaging refractometer is not simply the Fourier transform of the shadowgraphy image - they contain distinct and complementary sets of information about the rays exiting the plasma.

The distribution of line-integrated deflection angles measured by the imaging refractometer is directly related to line-integrated density gradients within the plasma:\cite{Hutchinson2002}
\begin{eqnarray}
        \phi = \frac{1}{2} \int \frac{\nabla n_e}{n_{cr}}dl 
\label{eq:syn_phi}
\end{eqnarray}
In these experiments we achieved an angular resolution of $\simeq 0.1$ mrad, and could measure angular deflections of $\pm 30$ mrad (limited by CCD size), giving a dynamic range of $\sim10^3$.\cite{Hare2021}

\subsection{An example of imaging refractometer data using a grid}

To better understand and experimentally validate the imaging refractometer response under real plasma conditions, we placed a high-transparency planar grid in the flow from an exploding wire array to generate a network of interacting oblique shocks.
The grid consisted of 16 evenly spaced copper rods with a diameter, D = 50 \normalfont{$\mu$}m, and a spacing, $\delta_0$ = 800 \normalfont{$\mu$}m, oriented parallel to the y-axis (which is the same orientation as the advected magnetic field). The shocks which are generated thus have the same spatial periodicity as the grid.
The periodicity of the flows in this experiment is similar to the sinusoidal density perturbation used in the ray-tracing simulations shown in Ref. \onlinecite{Hare2021}.

Fig. \ref{fig:grid_Al_shadow}(a) shows the side-on shadowgraphy image of the plasma structure formed by the interaction of the plasma flow with the grid at 404 ns after current start. The plasma flow from the left side of the image ($x<0$ mm) collides with the grid rods at x = 0 mm. In the post-grid region ($x>0$ mm), multiple oblique shocks emerge at x = 0.5 mm, expanding as the flow propagates to the right.
The oblique shocks are evenly spaced by $\delta = 0.8$ mm in the z-direction, corresponding to the grid wire spacing, and these shocks generate strong density gradients which deflect the probing laser beam. 
Further from the obstacle grid the strength of the shocks reduces due to the combined effect of the decreasing overall plasma density imposed by the current history and the radial divergence of the flows. 
The interaction of these shocks is visible in the shadowgraphy image at x = 0.5 mm, x = 1.8 mm, and x = 4 mm. Although it is not visible in the shadowgraphy image, an additional intersection occurs at x = 6.7 mm, as highlighted by the projected lines in the figure.
This interaction is visible in the refractometer image as a broadening of the deflection angle distribution in Fig. \ref{fig:grid_Al_shadow}(b), which we will discuss in more detail below.

Fig \ref{fig:grid_Al_shadow}(b) shows the signal from the imaging refractometer, which measures significant deflections of the probing laser beam. The distribution of deflection angles is symmetric about $\phi$ = 0 mrad, as expected by the symmetry of the obstacle, and the most striking features are the strong periodic modulations extended along the horizontal (spatial) axis of the image which will be discussed further below.
For now, we focus on the broad envelope of the deflection angle distribution, which begins very broad (up to 30 mrad) close to the grid, but decreases as the flows propagate further from the obstacle,  consistent with a weakening of the density gradients in the post-obstacle flow. At x > 10 mm, the signal on the detector is only slightly broader than the response function of the diagnostic, measured from a background image taken before the experiment. 

As shown in eq. \ref{eq:syn_phi}, stronger density gradients produce larger ray deflections, resulting in the broadening of the intensity distribution around its central value. Without any assumptions on the type of distribution, the standard deviation of the signal intensity represents a statistical description of angular deflections for a given spatial position, 

\begin{eqnarray}\label{eq:phi_std}
    \sigma_\phi = \sqrt{\sum_i \phi_i^2 I(\phi_i)}/\sum_i I(\phi_i)
\end{eqnarray}

where $I(\phi_i)$ is the intensity of the pixel corresponding to the deflection angle $\phi_i$.
Higher-order moments of the deflection angle distribution function, such as skewness and kurtosis, may also be used to analyse the distribution of angular deflections and possibly relate them to the spectrum of density fluctuations within a plasma. However, this paper will focus solely on the standard deviation and further statistical analysis will be explored in future papers.



Two examples of intensity distributions of angular deflections are shown in Fig. \ref{fig:grid_Al_shadow}(d) \& (e), taken at x = 4 mm and x = 6.7 mm respectively. In both cases, the intensity signal is much broader than the characteristic diagnostic response (green profile, $\sigma_\phi\approx 0.12$ mrad).
By extending the standard deviation analysis to angular deflection distributions at each spatial position x, the overall standard deviation profile, $\sigma_\phi$, can be reconstructed and is reported in Fig. \ref{fig:grid_Al_shadow}(c).

The locations of the oblique shock interactions are visible in Fig. \ref{fig:grid_Al_shadow}(c) as a sudden increase in $\sigma_\phi$ at x = 4 mm and x = 6.7 mm, corresponding to the locations seen in the shadowgraphy image. 
It is notable that the peaks in $\sigma_\phi$ occur after the corresponding crossing points measured from the shadowgraphy image. This is due to the fact that the locations of the shock intersections in the x-axis vary along the z-axis, consistent with the axial expansion of the flows into the vacuum regions above and below the arrays.
Although the third shock crossing is not visible in the shadowgraphy image, it can be clearly seen in imaging refractometer data at x = 6.7 mm, demonstrating the sensitivity of this diagnostic to very small deflection angles.

In this experiment, the density perturbations have the same periodicity as the grid, and we observe periodic modulations in the deflection angle distribution, spaced by $\Delta\phi \approx 1.2$ mrad [Fig. \ref{fig:grid_Al_shadow}(b) and Fig. \ref{fig:grid_Al_shadow}(d) and (e)]. 
These modulations have a striking visual similarity to the interference patterns caused by a periodic opaque structure or diffraction gratings. In this experiment, however, there is no opaque structure present and so this analogy is somewhat misleading. Instead, each shock acts as a refracting object, deflecting light rays at various angles. The periodicity of the shocks combined with the varying path differences caused by different refraction angles leads to constructive and destructive interference which is then captured by the imaging refractometer camera. The angular periodicity of the interference pattern can be related to the period of the shock system through:\cite{hecht2012optics}
\begin{eqnarray}
    \Delta \phi_n =  \frac{n \lambda_0}{\delta}  =  \frac{1.053 \; \mu m}{800 \; \mu m} \approx n \cdot 1.3 \; \mathrm{mrad} \; (n \in \mathbb{Z})
    \label{eq:shock_periodicity}
\end{eqnarray}
By substituting the known shock period along z imposed by the grid, $\delta$, and the laser wavelength, $\lambda_0$, we find a remarkable agreement between this simple theory and the measured $\Delta \phi \approx 1.2$ mrad, which demonstrates that these interference patterns in the imaging refractometer can be used to estimate the length scale of periodic density modulations in the plasma.
Having demonstrated the efficacy of the imaging refractometer in this experiment with clearly defined density modulations, we now turn this diagnostic towards experiments where the density modulations arise from the plasma itself, rather than being imposed by the obstacle.

\section{\label{sec:ExperimentalResults} Results from magnetised flow interactions with planar obstacles}

In this section, we discuss the collision of aluminium and tungsten plasma flows with a planar conducting obstacle (copper plate 35x35 mm, 500 um thickness). For both experiments, we present an analysis of refractometry images and corresponding shadowgraphy images, along with quantitative measurements of angular deflections, as we did for the case with the grid above. In the case of the tungsten wire arrays, additional measurements using end-on interferometry and optical Thomson scattering are also included.
    
\subsection{Aluminium plasma flows with a planar conducting obstacle}  

    \begin{figure*}
    \centering    
    \includegraphics[width = 1\linewidth]{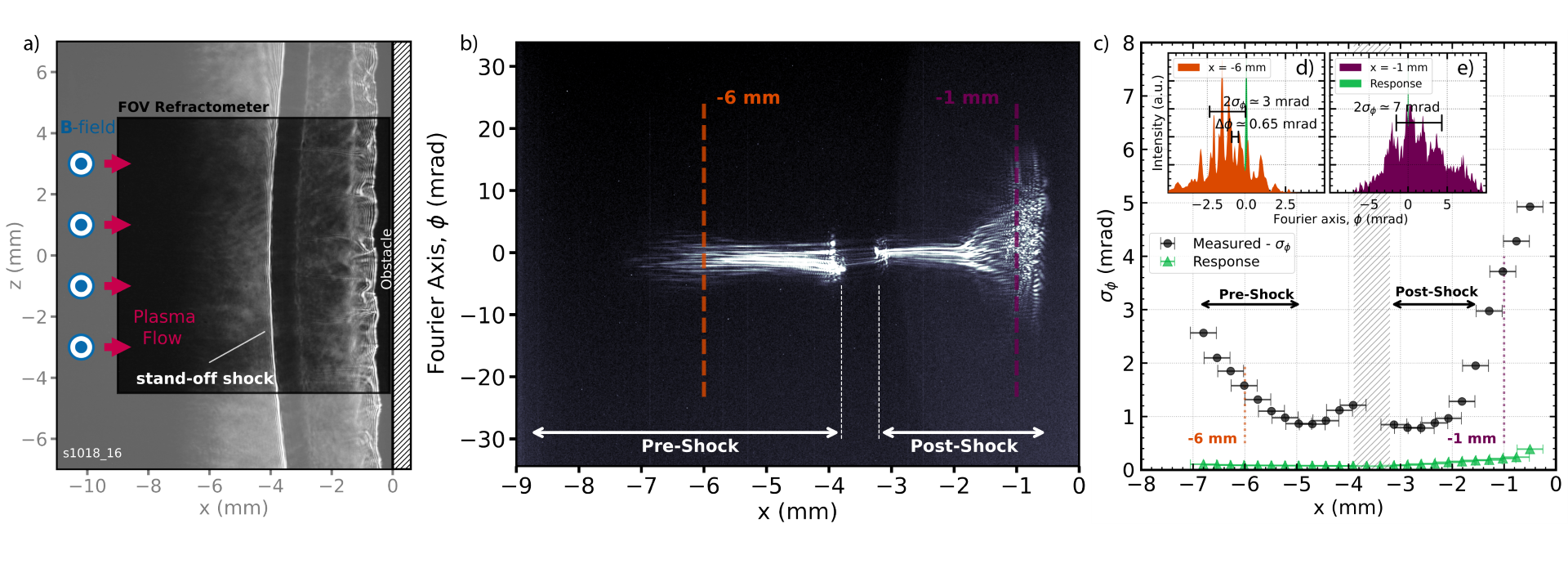}
    \caption{
        (a) 1053nm side-on shadowgraphy image of aluminium plasma flow colliding against a planar obstacle, taken at 391 ns after current start. The origin of the horizontal axis is assumed to be at the obstacle position which was placed at $\SI{10}{mm}$ from the exploding wire array. The stand-off shock is formed at x = -4 mm. Visible caustics are present at the beginning of the stagnation layer, at x = -1 mm from the planar wall. (b) Refractogram of the FOV region highlighted in the shadowgraphy image. (c) The measured standard deviation of the refractometry image signal, $\sigma_\phi$, taking $\Delta x = 0.4$ mm from a single experiment. Almost no signal is detected near the shock transition (masked region). (d), (e) Example of angular deflection distribution at x = -6 mm and x = -1 mm respectively. Broader angular deflections are observed in the regions closer to the obstacle. Peak displacements resulting from interference effects are spaced with $\Delta \phi = 0.65$ mrad, visible both upstream of the stand-off shock and just upstream of the obstacle.
    }

    \label{fig:planar_al_shadow}
\end{figure*}

Fig.\ref{fig:planar_al_shadow} shows side-on data from the aluminium plasma experiment at 391 ns after the start of the current pulse. In the shadowgraphy image of Fig. \ref{fig:planar_al_shadow}(a), the plasma flow moves from left to right and collides with the planar obstacle at x = 0 mm.
As in previous aluminium experiments, we observed a stand-off shock at $x = \SI{-4}{\mm}$ due to the pile-up of the magnetic field which is perpendicular to the flow direction, and a dense stagnated region near the obstacle surface, $x > \SI{-1}{\mm}$.\cite{Lebedev2014} 

In the pre-shock region, $\SI{-10}{mm}<x<\SI{-6}{mm}$, the shadowgraphy signal is strongly attenuated, which we attribute to the strong inverse bremsstrahlung absorption of the probing beam, though we expect that the plasma density is less than the critical density for this probing wavelength ($n_{cr} = 10^{21}$ cm$^{-3}$). 
The plasma density drops away from the array, allowing the beam to pass unattenuated to the camera, and in the region $\SI{-5}{mm}<x<\SI{-4}{mm}$ the shadowgraphy image shows a uniform intensity distribution corresponding to small density perturbations. 
A curved bright region is very clear around x = -4 mm, which corresponds to a strong jump in plasma density which we attribute to a stand-off shock caused by the pile-up of mass and magnetic field against the conducting planar obstacle. 
Next to this bright region, there is a corresponding dark region, from which the laser light has been refracted due to the density gradient. 
Further downstream the intensity is relatively uniform, and close to the obstacle ($\SI{-1}{mm}<x<\SI{-0.5}{mm}$) we observe large perturbations to the laser intensity, with caustic structures consistent with large density gradients. 
In the region $x>\SI{-0.5}{mm}$, very close to the obstacle, the laser beam is not captured by the camera due to a combination of refraction and absorption effects in the stagnated plasma on the obstacle surface.

The refractometry image in Fig. \ref{fig:planar_al_shadow}(b) was taken in the same experiment at the same time as the shadowgraphy image, with the field-of-view region highlighted in Fig. \ref{fig:planar_al_shadow}(a). As in the shadowgraphy image, we see the signal only for $x>\SI{-7}{\mm}$ due to strong absorption. Upstream of the stand-off shock, the refractometer signal shows a narrow distribution of deflection angles which becomes narrower close to the shock. In addition, the intensity distribution is clearly modulated along the vertical, angular axis, consistent with the presence of periodic density modulations known to be present in wire array ablation, which we will discuss later. 

At the shock transition, the strong density gradient causes a loss of signal due to shadowgraphy effects in the region $\SI{-3.9}{mm}<x<\SI{-3.2}{mm}$. In the post-shock region, the total signal width is narrower ($\approx 50\%$ lower) with signs of elongated coherent structures remaining from the upstream perturbation. Close to the obstacle the distribution of deflection angles becomes much broader, and the intensity modulations appear less coherent. Past the stagnation interface, at x > -1 mm, the overall angular deflection signal reaches the maximum width while the intensity distribution is dominated by small-scale intensity modulations.


Fig. \ref{fig:planar_al_shadow}(d) \& (e) shows two examples of angular deflection distributions in the pre-shock and post-shock regions. The width of the deflection angle distribution more than doubles from $2\sigma_\phi \approx \SI{3}{\milli\radian}$ at x = -6 mm, upstream of the shock, to $2\sigma_\phi \approx \SI{7}{\milli\radian}$ at x = -1 mm, downstream of the shock. In Fig. \ref{fig:planar_al_shadow}(c) we show the standard deviation of the angular deflection distributions $\sigma_\phi$ for each spatial position, averaged over $\Delta x = \SI{0.4}{\mm}$ from a single experiment. In the plasma upstream, the laser angular deflection gradually decreases away from the wire array, reaching the minimum of $\approx \SI{0.8}{\milli\radian}$ at x = -5 mm, before increasing again as it approaches the stand-off shock. In the post-shock region, a gradual increase in angular deflections of the laser beam is observed from $\approx \SI{0.8}{\milli\radian}$ up to $\approx \SI{5}{\milli\radian}$.

\subsection{Tungsten plasma flows with a planar conducting obstacle}

\begin{figure*}
    \centering    
    \includegraphics[width = 1\linewidth]{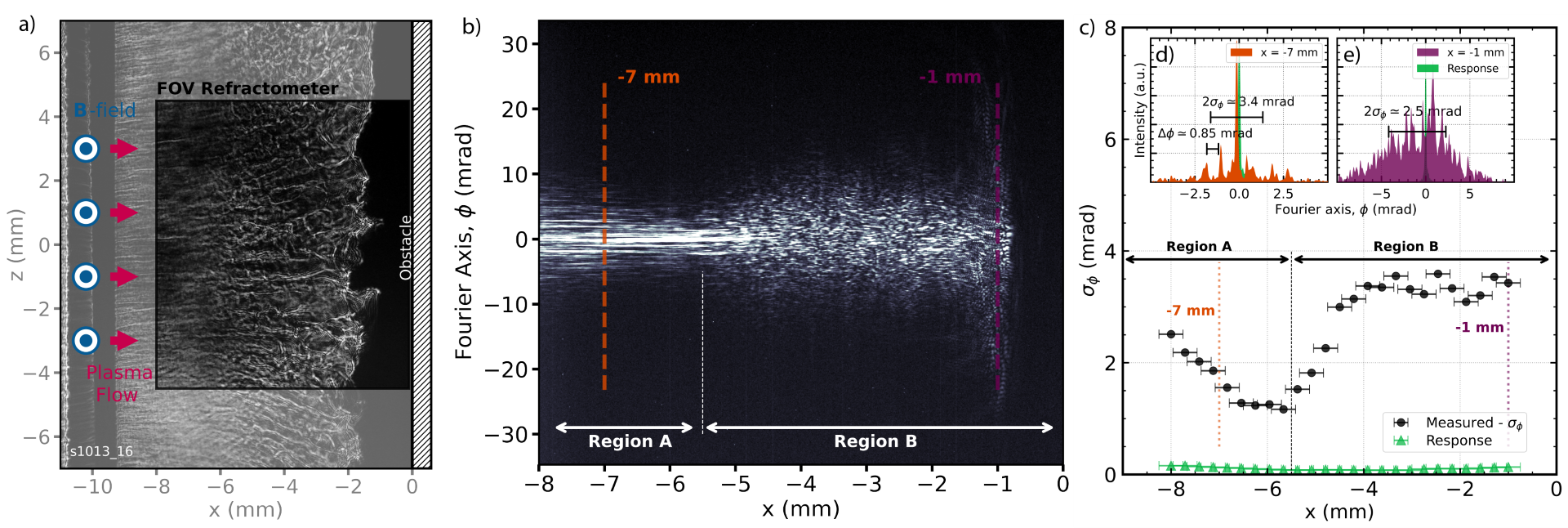}
     \caption{
        (a) 1053 nm side-on shadowgraphy image of tungsten plasma flow colliding against a planar obstacle at 440 ns after current start. The origin of the horizontal axis is at the obstacle position which was placed at 10 mm from the exploding wire array. The initial perturbations imposed by the wire ablation phase are visible in the upstream flow until x = -6 mm, while small-scale structures dominate the plasma flow up to the stagnation layer. (b) Refractometry image of the FOV region highlighted in the shadowgraphy image, which shows two distinct regions: Region A (corresponding to the flow from the wire array), and Region B (corresponding to the interaction of that flow with the planar obstacle). (c) The standard deviation of the refractometry image signal, $\sigma_\phi$, taking $\Delta x = 0.4$ mm from a single experiment. (d), (e) Example of angular deflection intensity profiles at x = -7 mm and x = -1 mm from the obstacle, respectively. After the transition from region A to region B, density gradients increase together with the angular broadening, $\sigma_\phi$. Peak displacements resulting from interference effects are equally spaced with $\Delta \phi \approx 0.85$ mrad, visible in the upstream flow.
    }
    \label{fig:planar_w_shadow}
\end{figure*}

We now present results from the same experiment, but using tungsten wires. Although both aluminium and tungsten plasmas have similar upstream mass density and flow velocity, side-on diagnostic images revealed a dramatically different behaviour. Fig. \ref{fig:planar_w_shadow}(a), shows a shadowgraphy image captured at 440 ns after current start, where again we have chosen the origin of the coordinate system such that the wires are located at x = -10 mm, and the obstacle surface at x = 0 mm. The wires are clearly visible, in contrast to the shadowgraphy image from the aluminium experiment shown in Fig. \ref{fig:planar_al_shadow}(a), which we attribute to the lower electron density in the tungsten plasma flows.

As in the aluminium experiments, we observe periodic intensity modulations in the deflection angle distribution, extended along the spatial direction up to x = -6 mm. 
In contrast to aluminium plasma, there is no well-defined bright region corresponding to a stand-off shock. Instead, at x > -6 mm the periodic intensity modulations are less pronounced, and smaller perturbations start to dominate the image. In the region between x = -6 mm and x = -2 mm, the intensity distribution appears fragmented with intensity variations on a wide range of spatial scales. 
Close to the obstacle (x > -2 mm), there is a rippled opaque region of stagnated plasma.

Fig. \ref{fig:planar_w_shadow}(b) shows the refractometer image. Two distinct regions can be identified which are also visible in the shadowgraphy data: Region A, from the wire array at x = -10 mm to x = -5.5 mm, and Region B from x = -5.5 mm to the obstacle surface at x = 0 mm. 
Similar to the aluminium case, region A shows a decrease in width of the distribution of deflection angles as the plasma flow expands outward from the array, modulated by the interference effects from the wire array ablation perturbations discussed below. In Region B, the distribution of deflection angles broadens, reaching $\approx 7$ mrad over a broad region. Close to the obstacle the distribution broadens even further and appears fragmented with small-scale fluctuations in x and $\phi$.

Two angular deflection distributions taken from Region A and Region B are shown in Fig. \ref{fig:planar_w_shadow}(d) \& (e). At x = -7 mm, the laser is mostly deflected by angles less than $\sigma_\phi \approx 1.7$ mrad, modulated by evenly spaced peaks separated by $\Delta \phi \approx 0.85$ mrad. In Region B at x = -1 mm, the intensity profile is much wider, $2\sigma_\phi \approx 7$ mrad with no obvious periodicity.

 Fig. \ref{fig:planar_w_shadow}(c) shows the measured standard deviation $\sigma_\phi$, which decreases to a minimum of $\sigma_\phi \approx 1$ mrad at x = -5.5 mm at the edge of region A. The transition between regions A and B is marked by a sharp increase in the width of the distribution of angular deflections, resulting in a maximum of $\sigma_\phi \approx 3.5$ mrad. For x > -4 mm, the angular deflection signal remains almost constant at around 3 mrad. At all positions in the refractometry image, the angular deflection is much wider than the diagnostic response (green line), with $\sigma_{\phi, signal}/\sigma_{\phi, response}\approx 200 $.
    
\subsubsection{Electron density and Thomson scattering measurements of the tungsten plasma} 

\begin{figure*}
    \centering
    \includegraphics[width = 1\linewidth]{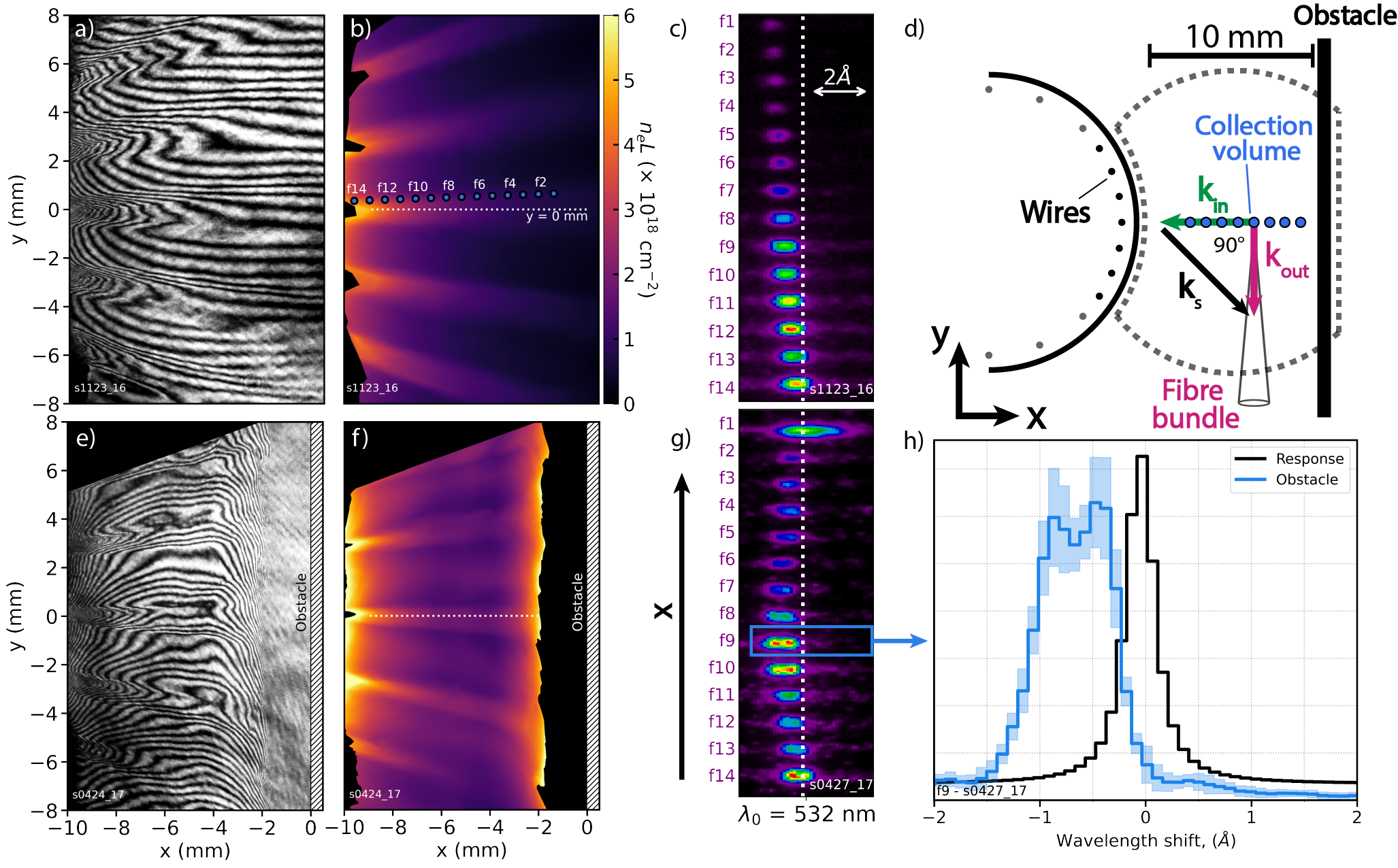}
    \caption{
    (a) Laser interferogram taken at 384 ns after current start, (b) processed end-on electron density map of the tungsten plasma flow and (c) Thomson scattering spectrogram, without any obstacle, in the same experiment, at 365 ns after current start. (d) Schematic illustration of Thomson scattering geometry. (e) Laser interferogram taken at 348 ns after current start, (f) processed end-on electron density map, (g) Thomson scattering spectrogram of the ion feature in the tungsten plasma flow interacting with a planar obstacle, in a different experiment with identical conditions, taken 324 ns after current start. (h) Example of Thomson scattering spectrum from fibre 9 of the spectrogram (x = -5.3 mm) in the presence of the obstacle. The shaded region represents the standard deviation of the signal, and the solid line is the average over one fibre in the vertical direction of the spectrogram, to boost the signal-to-noise ratio against self-emission and shot noise. The response of the spectrometer to the unshifted laser probe is  plotted for comparison.
    }
    %
    \label{fig:planar_w_if_ots}
\end{figure*}

Further measurements were performed in the tungsten plasma experiment using a combination of a spatially resolved Thomson scattering diagnostic and end-on laser interferometry. Thomson scattering and interferometry data for the aluminium experiment were presented in Refs. \onlinecite{Lebedev2014, Burdiak2017}.
Fig. \ref{fig:planar_w_if_ots} shows measurements of electron density and raw Thomson Scattering spectra in exploding wire-array experiments without and with the planar conducting obstacle (top and bottom row, respectively). The data were collected from three different experiments performed under similar conditions: without the obstacle, the interferometry and Thomson scattering measurements were made during the same experiment at 384 ns and 365 ns after current start, respectively, and with the obstacle the interferometry and Thomson scattering measurements were made during separate experiments at 348 ns and 324 ns after current start, respectively.
These times are all similar - the system evolves on timescales of 100 ns due to the current rise time of 240 ns.
The unprocessed interferograms are shown in Fig. \ref{fig:planar_w_if_ots}(a) \& (e). The interaction with the obstacle leads to a dense stagnation layer at x = -2 mm, and interference fringes are lost in this region as the probing beam is refracted out of the collection optics and lost.

\begin{figure}[H]
    \centering
    \includegraphics[width = 0.99\linewidth]{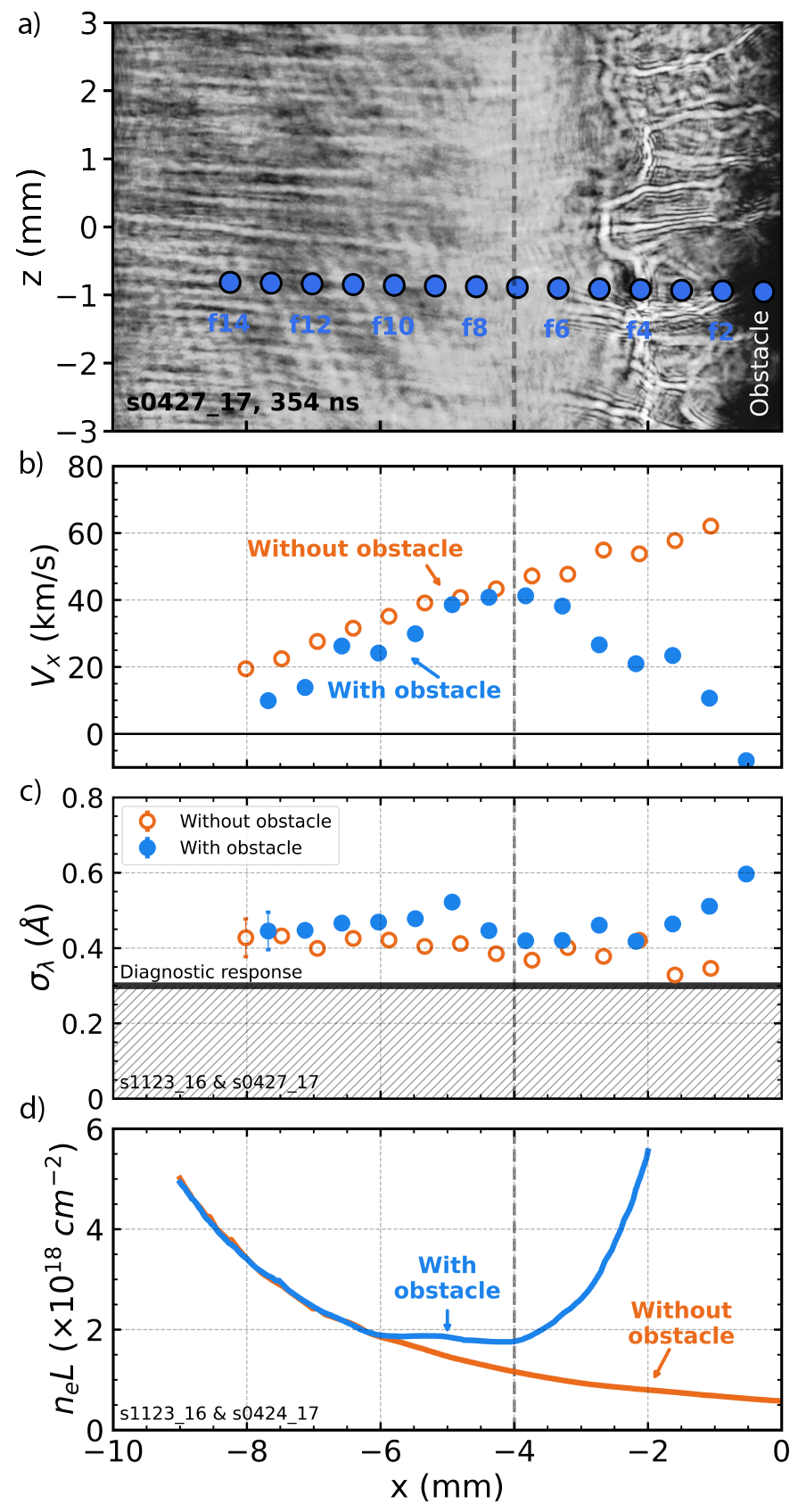}
    \caption{
    (a) Side-on shadowgraphy image of the tungsten plasma flow taken at 354 ns after current start using a 532 nm laser, with the location of the 14 TS collection volumes overlaid. (b) Velocity measurements of the plasma with and without the planar obstacle. The uncertainty in the velocity measurement is negligible ($\pm$ 2 km/s) resulting in error bars smaller than the size of the markers. (c) Broadening profile of the TS spectra. $\sigma_\lambda$ was calculated as one intensity-weighted standard deviation of the spectrum distribution. The error bars on the first two measurements are representative of the other data points and are estimated based on the uncertainties of the Thomson scattering spectra [Fig. \ref{fig:planar_w_if_ots}(h)]. (d) $n_e L$ profiles taken at y = 0 mm from the end-on interferometry data in Fig. \ref{fig:planar_w_if_ots}, at 348 ns after current start with the obstacle (blue line) and 36 ns later (384 ns after current start) without the obstacle (orange line).
    }
    \label{fig:planar_w_shadow_vel}
\end{figure}

Fig. \ref{fig:planar_w_if_ots}(b) \& (f) show the processed line-integrated electron density maps $n_e L$ in the xy-plane ($n_e$ can be estimated by assuming L = 22 mm, the array height). 
As the plasma flow expands radially into the vacuum the electron density decreases, reaching $n_e \approx 5 \times 10^{17}$ \text{cm}$^{-3}$ at 10 mm from the array. 
In the presence of the planar obstacle, plasma stagnates at the obstacle surface, forming a dense region that blocks the probing laser for $x > -2$ mm.
Lineouts of the electron density with and without the obstacle are shown in Fig. \ref{fig:planar_w_shadow_vel}(d).

The end-on electron density maps show little evidence of the small-scale density structures evident in the shadowgraphy and imaging refractometer images in Fig. \ref{fig:planar_w_shadow}, which were seen in the plasma from the side-on line of sight. 
The ability to resolve small-scale structures is limited partially by the spatial resolution of the electron density maps caused by interpolating data between interference fringes.
However, these line-integrated electron density maps are also consistent with density structures elongated in the y-direction along the magnetic field, which are narrower in the z-direction and randomly distributed, causing them to be washed out by the line integration. This will be discussed further below.

Thomson scattering measurements were taken at comparable times to the interferometry data, allowing us to make detailed local measurements of the plasma flow velocity. Fig. \ref{fig:planar_w_if_ots}(c) $\&$ (g) shows raw TS spectra with and without the planar conducting obstacle. 
The spectrograms display the spectrum from each fibre on the vertical axis, corresponding to a discrete spatial position in the plasma. The horizontal axis shows the spectrum for each fibre, spanning the ion feature of the Thomson scattering spectrum. These measurements were taken using the geometry illustrated in Fig. \ref{fig:planar_w_if_ots}(d), with the laser probing beam directed anti-parallel to the flow direction, and the scattered light observed from one direction at 90$^\circ$ to the probing laser. This gives a resultant scattering vector at 45$^\circ$ to the flow direction, and so the resultant Doppler shift in the spectrum will contain a component related to the radial flow velocity. An example of a TS spectrum collected from fibre 9 is shown in Fig. \ref{fig:planar_w_if_ots}(h) which shows significant Doppler broadening and a Doppler shift compared to the response function of the spectrometer, which was measured using the reflection of the probe beam from a metallic obstacle.

Results of the TS analysis are summarised in Fig. \ref{fig:planar_w_shadow_vel} together with a side-on shadowgraphy image at a comparable time to TS data for the obstacle case. At 350 ns after current start, the perturbed region in Fig. \ref{fig:planar_w_shadow_vel}(a) extends between $\SI{-4}{\mm} < x < \SI{0}{\mm}$, while the upstream remains dominated by the periodic density modulation discussed above. 
Fig. \ref{fig:planar_w_shadow_vel}(b) shows the spatially resolved measurements of flow velocity, $V_x$, with and without the planar conducting obstacle, calculated by measuring the Doppler shift and projecting the scattering vector onto $V_x$, assuming negligible $V_y$. We also calculate the broadening of the TS signal, $\sigma_\lambda$, by taking the standard deviation of the spectrum. Each data points correspond to one of the 14 TS collective volumes as indicated in Fig. \ref{fig:planar_w_if_ots}(b) and Fig. \ref{fig:planar_w_shadow_vel}(a).

In the absence of an obstacle [Fig. \ref{fig:planar_w_shadow_vel}(b)], we observe that the flow velocity shows a roughly linear increase with distance from the wire array, starting from the 20 km/s near the wire array at x = -8 mm and reaching a maximum velocity of 60 km/s at x = -1 mm. This acceleration has been seen in 3D MHD simulations of exploding wire arrays.\cite{Datta2022a} When the obstacle is present the flow velocity matches that without the obstacle for $x < -4$ mm, reaching 40 km/s. Beyond this point, the velocity profiles for the experiments with and without the obstacle begin to differ, with the flow gradually decelerating for the case with the obstacle.

Fig. \ref{fig:planar_w_shadow_vel}(c) shows the measured width of the Thomson scattering spectra as one intensity-weighted standard deviation of the spectrum distribution. When the obstacle is not present, the scattered signals are only 40\% wider than the diagnostic resolution, which is approximately \SI{0.3}{\angstrom}, and the width of the spectrum remains nearly constant for all the spatial positions measured. In the presence of the obstacle, the broadening is similar for the first few spatial positions with x<-6 mm. However, we observe a measurable increase in $\sigma_\lambda \approx \SI{0.5}{\angstrom}$ ahead of the perturbed region, at x = -5 mm. The spectra become narrower again before becoming much broader close to the obstacle, x > -2 mm, in the region which cannot be probed using the interferometer and in which we observed significant density perturbations using the side-on shadowgraphy and imaging refractometer.

The spectra measured are only slightly broader than the response function of the spectrometer, which makes it challenging to infer a plasma temperature from the Doppler broadening, as is usually done with Thomson scattering spectra. This is an issue for high atomic mass plasmas such as tungsten, as the broadening due to the splitting of the ion acoustic waves is proportional to $C_s\sim\sqrt{\bar{Z}T_e/m_i}$. In the experiment with the obstacle, we only see evidence of distinct ion acoustic peaks in fibres 9 and 10 [Fig. \ref{fig:planar_w_if_ots}(g)]. 
We estimate that $\bar{Z}T_e \approx 140$ eV (Ref. \onlinecite{Froula2011} eq 5.4.4) is necessary to produce these peaks, which represents a significant increase in the electron energy in these regions compared to $\bar{Z}T_e < 60$ eV in the upstream region (fibres 14-11) of this experiment, and in the experiment without an obstacle [Fig. \ref{fig:planar_w_if_ots}(c)].
For the regions without distinct ion acoustic peaks, either the plasma is too cold for the peaks to be well separated, or the density perturbations observed with the imaging refractometer give rise to associated velocity fluctuations within the scattering volume which broaden the TS spectra and obscure the peaks.
The spectral broadening is  particularly apparent for fibre 1 in Fig. \ref{fig:planar_w_if_ots}(g), which collects light from the volume closest to the obstacle - here $\sigma_\lambda$ corresponds to around 700 eV of thermal motion, though the partition between electrons and ions is unclear. In this region we also observe a Doppler shift in the opposite direction to the other fibres, suggesting flow away from the obstacle.

Fig. \ref{fig:planar_w_shadow_vel}(d) shows electron density profiles taken along the TS probing beam at y = 0 mm from the end-on interferometry data. The upstream densities for x < -6 mm in the experiments with and without the planar obstacle are very similar (the time difference is 36 ns), and the density decreases with distance from the array due to divergence and time-of-flight effects. Notably, the electron density in the presence of the obstacle appears to flatten in the region -6 mm < x < -4 mm, where the flow acceleration phase ends and the TS spectra initially become broad, even though there are no associated intensity fluctuations in the shadowgraphy image [Fig. \ref{fig:planar_w_shadow_vel}(a)]. The density profile in the deceleration region, x > -4 mm, increases inversely with the flow velocity, consistent with incompressible, sub-sonic flows.

\section{\label{sec:Discussion} Discussion}
In this section, we compare and contrast the experiments carried out with aluminium and tungsten wire arrays, and try to explain the radically different structures we observed. In both experiments, the supersonic plasma flow is perturbed by the presence of the planar conductive obstacle, forming a deceleration region which extends up to x = -4 mm into the upstream flow. In the aluminium experiment, this transition is characterised by a quasi-one-dimensional shock mediated by the pile-up of the magnetic field at the obstacle surface. In the tungsten experiment, there is no well-defined shock, and density perturbations dominate the deceleration region ahead of the obstacle. Despite these striking differences, end-on interferograms show an increase in plasma density associated with the decrease in velocity in the region x > -4 mm, suggesting that a shock may still form in tungsten plasma. The presence of large density perturbations obscures the typical quasi-one-dimensional shock profile, leading to a complex, three-dimensional shock structure, similar to that hypothesised to exist in astrophysical accretion flows.\cite{Orlando2010, Reale2013, Matsakos2013}

The plasma parameters for the experiments with aluminium and tungsten wires are summarised in Table \ref{tab:plasma_params}, which serve as the basis for the discussion which follows. These parameters were taken just upstream of the interaction region, at x = -5 mm.

\begin{table}[htb]
    \centering
    \begin{tabular}{ @{\extracolsep{\fill}} lccc}
        \hline
        Quantity &  Symbol & Al & W \\ 
        \hline\hline
        Flow velocity (km/s)                                & V              & 50             & 40           \\
        Electron density ($\times 10^{18}$ cm$^{-3}$)       & $n_e$          & 4              & 0.9             \\ 
        Ion temperature (eV)                                & $T_i$          & 12             & < 10          \\ 
        Electron temperature (eV)                           & $T_e$          & 12             & < 10        \\ 
        Magnetic field (T)                                  & B              & 6            & 4            \\ 
        Average ionization                                  & $\bar{Z}$      & 3.5            & <6          \\ 
        \hline
        Mass density ($\mu$g/cm$^{3}$)                      & $\rho$         & 51             & 45         \\
        Ion-ion mean free path ($\mu$m)                     & $\lambda_{ii}$ & 4              & 5   \\ 
        Ion sound speed (km/s)                              & $c_s$          & 14             & 6         \\
        Alfv\'en speed (km/s)                               & $V_A$          & 24             & 17        \\
        \hline
        
        Sonic Mach Number                                   & $M_S$          & 3.6            & 6.6          \\ 
        Alfv\'en Mach Number                                & $M_A$          & 2.1            & 2.4          \\
        Reynolds Number                                     & $Re$           & $1\times10^{6}$  & $3\times10^{6}$ \\ 
        Magnetic Reynolds Number                            & $Re_M$         & 9             & 3          \\
        Dynamic beta                                        & $\beta_{dyn}$  & 9             & 11      \\
        Thermal beta                                        & $\beta_{th}$   & 0.7            & 0.3        \\
        \hline\hline        
    \end{tabular}
    \caption{Typical plasma flow parameters for aluminium and tungsten inverse wire arrays, from around 350 ns after current start and x = 5 mm downstream of the wire arrays (and hence 5 mm upstream of the obstacle). For aluminium arrays, data is taken from Ref. \onlinecite{Russell2022a} for $V$ [Fig. 3(c)], and Ref. \onlinecite{Burdiak2017} for $n_e$ [Fig. 4], $T_e$ and $T_i$ [Fig. 6(c)], $B$ [Fig. 2(b)], and $\bar{Z}$ using the atomic code SpK. \cite{Crilly2022} For tungsten arrays, the data is taken from this paper: $V$ is from Fig. \ref{fig:planar_w_shadow_vel}(b), $n_e$ is taken from Fig. \ref{fig:planar_w_shadow_vel}(d), $T_e$ and $T_i$ are inferred from the minimal broadening in Fig. \ref{fig:planar_w_shadow_vel}(c), $B$ from Faraday rotation data (not shown), and $\bar{Z}$ is computed using FLYCHK code\cite{Chung2005} assuming $T_e$ = 10 eV. $\lambda_{ii}$ is calculated using the thermal velocity of the ions. To calculate $Re$ and $Re_M$ we use $L = 5$ mm as a representative length scale for the pile-up of the magnetic field or the slowing of the bulk flow.}
    \label{tab:plasma_params}
\end{table}

\subsection{Modulation of the electron density in the plasma flows}
In the experiments with a planar obstacle, there is no grid to impose an axially periodic density perturbation. However, we still observe similar periodic structures in the distribution of deflection angles, suggesting that there is also a periodic density perturbation in the outflow from the wires. We attribute this to a quasi-periodic modulation of the ablation from wires in a wire array, arising from shear-flow-stabilised $m=0$ instability which has been extensively discussed in the literature of imploding and exploding wire arrays.\cite{Lebedev2001, Chittenden2008a, Harvey-Thompson2009} The wavelength of these perturbations is strongly dependent on the size of the wire cores, which for low-Z elements, such as Al, is observed to be larger than for high-Z materials, like W.

In a similar manner to the case with the grid, we can use the estimated angular periodicity, $\Delta \phi \approx 0.65 $ mrad, of these modulations [as illustrated in Fig. \ref{fig:grid_Al_shadow}(d)] to calculate a modulation wavelength for the aluminium experiments as $\approx 1.6$ mm. We note again that these modulations are invisible in the shadowgraphy diagnostic at this time after current start, but the sensitivity of the imaging refractometer reveals these small density perturbations.
In the experiments using tungsten wires, we observed the periodic modulation in the distribution of deflection angles as $\Delta \phi \approx 0.85$ mrad. This angular periodicity corresponds to a shorter wavelength ($\approx 1.2$ mm) than we observed above for aluminium plasmas, likely due to the smaller wire core size observed for tungsten wire arrays.
For both Al and W plasmas, the wavelength we measure using the imaging refractometer is longer than the ``natural mode'' reported in the literature ($\approx 0.5 $ mm for Al, $\approx 0.25 $ mm for W, from Ref. \onlinecite{Lebedev2001} at earlier times in the ablation process), but it is very close to the wavelength at which the global implosion instability develops at the end of the ablation phase.\cite{Lebedev2001}

\subsection{Spatial scale of the density perturbations in tungsten flows}
We now turn our attention to the small-scale density perturbations observed in side-on shadowgraphy and the imaging refractometer in the experiments with tungsten plasma flows, as shown in Fig. \ref{fig:planar_w_shadow}. The deflection angle $\Delta \phi$ of a ray from an electron density gradient is given by:

\begin{eqnarray}
    \Delta \phi = \frac{1}{2} \frac{\nabla n_e}{n_{cr}} \delta y
    \label{eq:syn_phi_discrete}
\end{eqnarray}

where \( \delta y \) is the length scale along the probing direction, and \( \nabla n_e = \Delta n_e/\delta z \) and the \(n_{cr}\) are the electron density gradient and the critical electron density respectively. We note that eq. \ref{eq:syn_phi_discrete} is simply the discrete version of eq. \ref{eq:syn_phi}.\cite{Hutchinson2002} We can rewrite the electron density change $\Delta n_e$, as a fraction $\epsilon = \Delta n_e / n_e \leq 1$ of the total electron density, giving:
\begin{equation}
    \Delta \phi \approx \epsilon \frac{n_e \delta y}{2 n_{cr} \delta z}
    \label{eq:single_density_gradient}
\end{equation}

This is the deflection from a single density gradient present along the probing path. Based on our experimental data, we assume that a ray will encounter many density gradients as it traverses the plasma, which motivates the use of a random walk model. 
Assuming random scattering, the total angular broadening of the laser beam along the y direction is the result of the accumulation of small deflection angles $\Delta \phi$ caused by the presence of small perturbations with spatial scales $\delta z$, along the entire plasma length, $L_y$:
\begin{equation}
    \sigma_\phi^2 = N_{steps} \Delta \phi^2 = \frac{L_y}{\delta y} \cdot \Delta \phi^2
    \label{eq:random_walk}
\end{equation}

If we substitute (\ref{eq:single_density_gradient}) into the random walk relation (\ref{eq:random_walk}), we can estimate the angular broadening $\sigma_\phi$ due to perturbations of amplitude $\epsilon \leq 1$:
\begin{equation}
    \sigma_\phi^2 = \frac{L_y}{\delta y}\frac{n_e^2\delta y^2\epsilon^2}{4 n_{cr}^2 \delta z^2}
\end{equation}

If we now assume that on small scales the density perturbations are isotropic with $\delta y = \delta z$,\footnote{In principle we can use information from other diagnostics to estimate the ratio of $\delta y/\delta z$ and use this to adjust the estimates which follow. For example, our end-on interferometry in Fig. \ref{fig:planar_w_if_zoom} shows $\delta y/\delta z \approx 1/5$. However, this level of detail is not necessary for the purposes of this order-of-magnitude estimate.} the spatial scale of density perturbations required to produce a given angular broadening is:

\begin{equation}
    \delta_y = L_y \frac{n_e^2}{4 n_{cr}^2 \sigma_\phi^2}\epsilon^2 
    \label{eq:perturbation_scale}
\end{equation}

In our experiments with tungsten, we measured angular spread $\sigma_\phi \approx 3.5$ mrad using an infrared laser probe (1053 nm, $n_{cr} \approx 10^{21}$ $\text{cm}^{-3}$), and assuming an average electron density $n_e \approx 1\times10^{18}$ $\text{cm}^{-3}$ and a total length $L_y = 10$ mm, equation (\ref{eq:perturbation_scale}) gives a spatial scale of approximately \(\delta_y \approx 200\) \text{$\mu$}m for the largest possible perturbation (\(\epsilon = 1\)). Any smaller perturbations, $\epsilon < 1$, would have required structures with spatial scales smaller by a factor of $\epsilon^2$ to produce the same total deflection $\sigma_\phi$, $\delta_y \approx 200 \cdot \epsilon^2 (\mu m)$. For example, for $\epsilon = 0.1$, the spatial scale of perturbations would be $\delta y \approx 2$ $\mu$m.
We note that even the largest scale, $\delta_y\approx 200$ $\mu$m is much smaller than the global length scale on which the velocity and density change [Fig. \ref{fig:planar_w_shadow_vel}], and is on the order of the Thomson scattering volume size and of the intensity variations seen in the shadowgraphy images [Fig. \ref{fig:planar_w_shadow}(a) and Fig. \ref{fig:planar_w_shadow_vel}(a)].

\subsection{Structure of the density perturbations observed in tungsten flows}

We observe the density perturbations very clearly in the side-on shadowgraphy [Fig. \ref{fig:planar_w_shadow}(a) and Fig. \ref{fig:planar_w_shadow_vel}(a)] and imaging refractometer [Fig. \ref{fig:planar_w_shadow}(b)], but the plasma appears very uniform when probed end-on using interferometry [Fig. \ref{fig:planar_w_if_ots}(e) and (f)]. Although interferometry has a lower spatial resolution due to the interference fringes, we still observed the density perturbations when using side-on interferometry (not shown) which suggests that the spatial resolution alone is not responsible for this discrepancy. This anisotropy in the observed density fluctuations suggests that they are elongated along the y-direction, and randomly distributed in the xz-plane, such that the line integration in the side-on y-direction reveals their presence, but in the end-on z-direction, they are washed out.

The presence of density perturbations elongated along the magnetic field lines raises the question of what role the magnetic field may play in the pressure balance necessary to support these small-scale density perturbations. We can calculate the resistive length scale using the inferred plasma parameters and the Spitzer-Braginskii resistivity as $L_\eta \approx$ 1.3 mm. This implies that magnetic diffusion dominates over advection at length scales smaller than $L_\eta$, and so the magnetic field is homogeneous on scales below $L_\eta$. As we inferred the existence of density perturbations at least an order of an order-of-magnitude smaller than this using the imaging refractometer, we conclude that the magnetic field is smooth and therefore not involved in these density perturbations. We note that the magnetic field can still pile-up against the conducting obstacle on larger length scales, and so the magnetic pressure may be responsible for decelerating the plasma flow at $x>\SI{-4}{\mm}$ as seen in Fig. \ref{fig:planar_w_shadow_vel}(b).

\subsection{Cause of the density perturbations}

\subsubsection{Electro-thermal instability}

The first instability we consider is the electro-thermal instability (ETI), which relies on a change in resistivity, $\eta$, with temperature. This instability has been observed in gas-puff z-pinches\cite{Ryutov2000b} and has been invoked to explain perturbations on imploding metal liners.\cite{Peterson2012, Awe2013} In these systems, the ETI grows much faster than the magnetised Rayleigh-Taylor instability, and so may be present before there has been a significant acceleration of the bulk plasma or metal.

In a plasma where the Spitzer-Braginskii resistivity $\eta\propto T_e^{-3/2}$ decreases with temperature, this instability takes the form of filaments in the direction of the current: the local Ohmic heating causes an increase in temperature, a drop in resistivity and hence a more attractive pass for current flow, which in turn leads to increased heating and further concentration of the current. In a metal where $\eta$ increases with $T$, the opposite is true: the localised Ohmic heating raises the resistivity, and the current diverts around the hotter regions, leading to perturbations perpendicular to the current flow direction in the form of striations.\cite{Ryutov2000b}

There are two reasons why the ETI is unlikely to be responsible for the density perturbations we observe. Firstly, the length scale for magnetic field perturbations (and hence the associated current density perturbations) is constrained to be  larger than the magnetic diffusive length scale $L_\eta\approx1.3$ mm, as discussed above. As current perturbations cannot be on length scales much smaller than this, the ETI is also ineffective at these small length scales, and so it cannot create anti-correlated density perturbations on small length scales. 

Secondly, in our shadowgraphy data [Fig. \ref{fig:planar_w_shadow}(a)] we observe structures which are predominantly striations with $k\parallel z$, parallel to the direction of current flow, rather than filaments with $k\parallel x$ perpendicular to the direction of the current flow. For these perturbations to be caused by the ETI, would require $d\eta/d T_e>0$, as in a metal, rather than the plasma we have. To investigate whether the ionization state $\bar{Z}(T_e)$ could lead to $d\eta/d T_e>0$, we calculated $\eta(T_e)$ using ionization tables from FLYCHK,\cite{Chung2005} and found no such region. As such, we conclude that the slope of the resistivity with temperature does not enable ETI striations to form in this plasma. Along with the permitted length scales due to magnetic diffusion, this rules out the ETI as an explanation for our observations.

\subsubsection{Kinetic effects}

One source of instabilities in plasma is kinetic effects where the plasma has free energy associated with a non-Maxwellian distribution, and the instability forms as the plasma relaxes to a Maxwellian. In a similar experiment to ours using aluminium wire arrays and a planar conducting obstacle, Ref. \onlinecite{Lebedev2014} saw evidence for the reflected ions in the Thomson scattering spectra at before 180 ns after current start, as the shock forms, and the reflected ions had velocities comparable to the bulk plasma flow (100 km/s), but in the opposite direction. These reflected ions could, in principle, trigger an instability such as the two-stream instability which could give rise to density perturbations, and this effect may be more pronounced in tungsten plasmas where the heavier ion mass gives rise to a longer ion-ion mean free path.
Evidence for interpenetration of counter-streaming flows has been seen in experiments with imploding tungsten wire arrays.\cite{Swadling2016}

However, Ref. \onlinecite{Lebedev2014} looked at early times than in our current experiments, when the density was much lower and hence the ion-ion mean free path was much longer. Ref. \onlinecite{Lebedev2014} saw no evidence for reflected ions at later times when the density is higher, such as at around 350 ns after current start when our observations are made, and we see no evidence for reflected ions in the Thomson scattering data obtained with tungsten wire arrays. This is consistent with the very short mean free paths that we calculate from our experimental parameters, as shown in Table \ref{tab:plasma_params}, in contrast to imploding wire array experiments where counter-streaming flows give a larger relative velocity and hence a longer ion-ion mean free path. Given the lack of evidence for reflected ions and the short mean free paths, we do not expect kinetic effects to play a significant role in the formation of the observed density perturbations.

\subsubsection{Radiative cooling}

Radiative cooling instabilities are well known in the astrophysical literature, following the analysis of Field. 
\cite{Field1965} 
Localised cooling leads to a drop in the plasma temperature, and hence a corresponding drop in the plasma thermal pressure. This leads to compression of the cooler volume, raising the density and therefore the cooling rate. This is a runaway process, but it is stabilised at small length scales by thermal conduction, and at large length scales by hydrodynamic motion. This instability has been attributed to the density perturbations seen in pulsed-power-driven colliding jet experiments.\cite{Suzuki-Vidal2015}

\begin{table}[H]
    \centering
    \begin{tabular}{ @{\extracolsep{\fill}} lcc}
        \hline
        Quantity, Symbol (Units) & Al & W \\ 
        \hline\hline
        Cooling function, $\Lambda$ ($10^{-22}$ erg cm$^3$/s)       & 9.8     & 210          \\
        Radiated power density, $P_{rad}$ ($10^{12}$ erg/cm$^3$s)   & 280     & 3400             \\ 
        Thermal energy density, $U_{th}$  ($10^{6}$ erg/cm$^3$)     & 31      & 28    \\ 
        Cooling time, $\tau_{cool}$ (ns)                            & 110     & 8        \\ 
        \hline
        Electron magnetisation, $\omega_{ce}\tau_e$                  & 0.7    & 0.3            \\ 
        Dimensionless thermal conductivity, $\kappa^c_\parallel [\kappa^c_\perp]$  & 6.1 [1.2] & 8.2 [2.8]  \\ 
        Thermal diffusivity, $\chi_\parallel [\chi_\perp]$   ($10^3$cm$^2$/s) & 75 [15]    & 67 [22]   \\
        \hline
        Field's length, $\lambda_{Field, \parallel} [\lambda_{Field, \perp}]$  ($\mu$m)  & 910 [410]        & 230 [140]   \\ 
        Isobaric length, $\lambda_{iso}$ ($\mu$m)                & 1400             & 50         \\
        \hline\hline        
    \end{tabular}
    \caption{Parameters for estimating the relevant time and length scales of the cooling instabilities in the plasma. The relevant plasma parameters are taken from Table \ref{tab:plasma_params}. The cooling functions are taken from FLYCHK,\cite{Chung2005, flychk-2023} using $n_e = 10^{18}$ cm$^{-3}$ and $T_e = 10$ eV for both Al and W. The dimensionless thermal conductivities ($\kappa^c_\parallel, \kappa^c_\perp$) are obtained using the tabulated coefficients in Ref. \onlinecite{Epperlein1986}.}
    \label{tab:cooling_params}
\end{table}
We consider this instability for both aluminium and tungsten wire arrays, using the parameters shown in Table \ref{tab:cooling_params}.
The cooling function is taken from tabulated values calculated using FLYCHK\cite{Chung2005, flychk-2023}, which are relatively sparse in density and temperature (every decade in cm$^{-3}$ for density, every 3-5 eV in the relevant temperature range), and for any atomic code there are significant uncertainties in the cooling function for high Z elements such as tungsten due to the vast number of electron energy levels involved. Therefore, we present the following analysis as indicative only, and we would require validated, high-resolution cooling functions to make a quantitative comparison between experiments and theory.

From the cooling function, we can calculate the radiated power density, $P_{rad} = n_i n_e \Lambda(n_i, T)$. The cooling time is defined as the ratio of the thermal energy density to the radiated power density, $\tau_{cool} = U_{th}/P_{rad}$. This timescale fundamentally sets the growth rate of the instability, and we see that it is much longer for Al (110 ns) than for W (8 ns). This suggests that cooling instabilities will grow much more rapidly for the tungsten case, consistent with the presence of small-scale density perturbations in experiments with tungsten wire arrays.

Formal analysis of the radiative cooling instability in Ref. \onlinecite{Field1965} yields a cubic equation whose three roots are modes of the instability. The stability of these modes depends on the derivatives of the radiative power density with respect to density and temperature. In the limit where the cooling time is much longer than the hydrodynamic time ($\tau_{hydro}/\tau_{cool}\ll1$ with $\tau_{hydro} = L/C_s$), the cubic equation reduces to a single solution, the stability of which depends on the derivative of the radiated power density with respect to temperature, and the thermal diffusivity. This limit is valid at length scales below the isobaric length, $\lambda_{iso} \sim C_s \tau_{cool}$, which represents the largest size that a pressure perturbation can grow to by radiative cooling. For Al, we find $\lambda_{iso} =1400$ $\mu$m, and for W it is much shorter, $\lambda_{iso} =50$ $\mu$m. At small scales, thermal conduction can balance radiative loss, stabilising the mode. The length scale at which this occurs is known as the Field's length, $\lambda_{Field} \sim (\chi \tau_{cool})^{1/2}$.\cite{Field1965, Pringle2007}

To calculate the Field's length, we need the thermal diffusivity. We begin by recognising that the presence of the magnetic field in our experiment reduces electron heat transport perpendicular to the magnetic field, giving rise to an anisotropic thermal conductivity. In this experiment the ions are unmagnetised, and so the ion heat transport is smaller than the electron heat transport by a factor of $(m_e/m_i)^{1/2}\sim 10^{-3}$, and hence we ignore the ion heat transport in the following analysis. For the representative parameters given in Table \ref{tab:plasma_params}, we calculate the Hall parameter $\omega_{ce}\tau_e$ as $\lesssim 1$, where $\omega_{ce} = eB/m_e$ and $\tau_e$ is the electron-ion collision time given by, for example, Ref. \onlinecite{Braginskii1965} eq. 2.5e. Despite this modest Hall parameter, we find the dimensionless perpendicular electron thermal heat conduction is smaller by a factor of 3 compared to the parallel value, using the tabulated values from Ref. \onlinecite{Epperlein1986}. We calculate the thermal conductivity, $\kappa = \kappa^c k_B n_e T_e \tau_e/m_e$ (Ref. \onlinecite{Braginskii1965} eq. 2.12) and the thermal diffusivity, $\chi =\kappa/n_e$, for both the parallel ($\parallel$) and perpendicular ($\perp$) directions. This gives similar values of $\chi_\parallel$ and $\chi_\perp$ for Al and W, with $\chi_\parallel > \chi_\perp$, as expected due to the magnetic field.

Finally, we can calculate the Field's length, the length scale of the smallest perturbation for a radiative cooling instability. For Al, this is 800 $\mu$m along the magnetic field and 360 $\mu$m perpendicular to the field. For W, we have an issue: the Field's length is derived assuming that the isobaric length, $\lambda_{iso}$, is large. However, we have calculated $\lambda_{iso} = 50$ $\mu$m, smaller than the estimated parallel Field's length $\lambda_{Field, \parallel} = 220$ $\mu$m. To resolve this contradiction would require us to consider the solutions to the cubic equation without taking the limit of $\omega\ll k c_s$, and to find the smallest length scale for which an unstable mode exists - this would be the new, generalised Field's length. However, for the uncertainties in the plasma parameters for our experimental data, and given the limitations in the currently available cooling curves, we instead state that we expect density perturbations on the order of 100 $\mu$m, elongated by a factor of two along the magnetic field lines due to the suppression of the thermal conductivity from partially magnetised electrons.

\begin{figure}
    \centering
    \includegraphics[width = 0.951\linewidth]{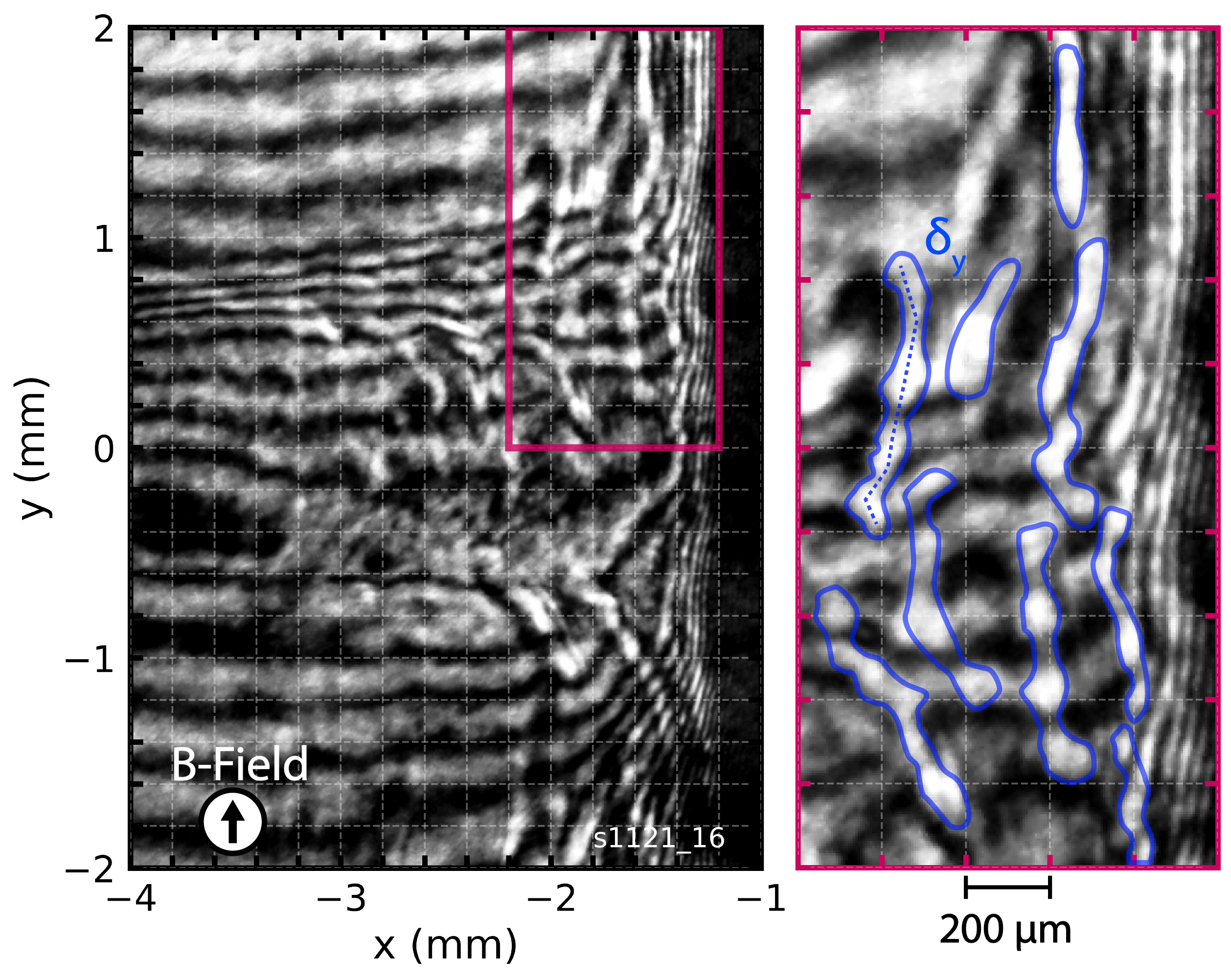}
    \caption{
    A section of a raw end-on interferogram taken at 391 ns after current start in an experiment with tungsten wires. The section shows a region close to the wall in which the brightness of the interference fringes is strongly modulated, suggesting shadowgraphy effects related to localised structures with large electron density gradients. The zoomed-in region shown on the right is overlaid with blue lines to highlight some of these structures, which are elongated in the y direction (along the magnetic field).
    }
    \label{fig:planar_w_if_zoom}
\end{figure}

The calculated perpendicular length scale ($\delta_x$, $\delta_z$) of these perturbations is in agreement with the length scales measured using our side-on shadowgraphy and imaging refractometer diagnostics, where we observed intensity fluctuations and deflection angle distributions consistent with density fluctuations with length scales $<200$ $\mu$m. In our end-on interferometry, the parallel length scale, $\delta_y$ is on the order of or smaller than the fringe spacing, and so it is difficult to resolve. However, careful examination of the raw interferograms shows some structure in the post-shock stagnation region, which we highlight in Fig. \ref{fig:planar_w_if_zoom}. This shows a section of an interferogram taken using a 532 nm laser at 391 ns after current start, in a region close to the obstacle and centred on one of the ablation streams. The brightness of the interference fringes is modulated on small scales, and we attribute this to shadowgraphy effects from the density gradients caused by the radiative cooling instability. To guide the eye, several of these intensity modulations have been highlighted in the further zoomed-in image, and it can be seen that these regions have a width of $\sim 100$ $\mu$m, and a length of $\sim 500$ $\mu$m.

Simulating these experiments requires validated, high-resolution cooling curves which have regions in which the radiative cooling function decreases with temperature,\cite{Pringle2007} and so this cooling instability will not appear in simulations using simple free-free (bremsstrahlung and recombination) radiation models. The lack of accurate cooling curves for high-Z elements, such as tungsten, limits our ability to simulate this experimental setup at this time. This also prevents us from calculating the threshold and growth rate for the thermal cooling instability, as it depends sensitively on $d \log \Lambda/d\log T$\cite{Suzuki-Vidal2015, Pringle2007} and is therefore difficult to evaluate for the sparsely-evaluated and unvalidated cooling functions we have access to for tungsten.

\section{Conclusions\label{sec:Conclusion}}

In this paper we have studied the presence of small-scale structures in the post-shock region of magnetised super-sonic plasma flows interacting with planar obstacles, using a range of diagnostics including the newly developed imaging refractometer.
We tested this imaging refractometer using an aluminium wire array and a periodic planar grid obstacle, and we observed the formation of a series of periodic, interacting oblique bow-shocks in the plasma flow downstream of the grid. These bow shocks provided the periodic density gradients, similar to the test problem reported in Ref. \onlinecite{Hare2021}, which led to a broad distribution of deflection angles modulated by interference effects.

We then turned our attention to the interaction of plasma flows with a planar conducting obstacle for aluminium and tungsten wire arrays. For both aluminium and tungsten arrays, we observed periodic modulations in the plasma flow using shadowgraphy and the imaging refractometer, and the periodicity inferred from the later diagnostic is in agreement with the literature on this shear-flow-stabilised $m= 0$ instability in wire-array z-pinches.\cite{Lebedev2001, Chittenden2008a, Harvey-Thompson2009}

For the aluminium wires, the plasma flow interacted with the planar obstacle to form a reverse shock, consistent with earlier work.\cite{Lebedev2014} In the post-shock region, we saw a reduction in the deflection angles in the imaging refractometer, corresponding to a smoother post-shock plasma. In contrast, with tungsten wire arrays we did not see a quasi-one-dimensional shock, and instead we observed a steady increase in the deflection angles as the plasma approached the obstacle, along with the presence of small-scale density perturbations. These perturbations were confirmed by shadowgraphy, and analysis of the deflection angles suggests a density perturbation length scale on the order of 100 $\mu$m. It is important to highlight that the absence of any sharp discontinuities in the tungsten experiment does not necessarily mean that a shock is not formed. The three-dimensional nature of the system and the presence of density structures ahead of the obstacle may obscure the typical quasi-one-dimensional shock profile, leading to a more complex shock structure, and this phenomenon is discussed extensively in the astrophysical literature.\cite{Orlando2010, Reale2013, Matsakos2013} Thomson scattering measurements show that the plasma begins to decelerate in the same region where the density perturbations increase and the Thomson scattering spectrum shows significant broadening in this region, consistent with velocity fluctuations on length scales smaller than the Thomson scattering collection volume, 200 $\mu$m. Associated with this decrease in velocity, end-on interferograms show an increase in plasma density which are both key signatures of a shock. In addition, a close look at the end-on interferograms reveals the presence of shadowgraphy effects consistent with density structures elongated along the direction of the magnetic field in the plasma.

To explain the formation of these density perturbations, we considered three mechanisms. Firstly, we looked at the electro-thermal instability that can create density perturbations with structures whose orientation with respect to the current direction depends on whether the resistivity of the material increases (metals) or decreases (plasmas) with temperature. However, the long magnetic diffusion length scale and the structure of the observed instability rule out the development of the small-scale current perturbations necessary for this instability to grow. Secondly, we looked at kinetic effects, in which reflected ions could give rise to a two-stream instability. We rejected this mechanism due to the short mean-free-paths of the ions in these relatively cold and dense plasmas, as well as the absence of any signatures of reflected ions in our Thomson scattering data.

Thirdly, we looked at radiative cooling instabilities. These occur due to a feedback loop which causes localised increased cooling as the local density increases to compensate for a loss of temperature. These cooling instabilities have length scales between the propagation of sound waves during a cooling time (the isobaric length, $\lambda_{iso}$), and the length at which thermal conduction smooths out temperature perturbations (Field's length, $\lambda_{Field}$), and they grow on a timescale on the order of the cooling time. For aluminium, this timescale is long (above 100 ns) compared to other timescales in the system, suggesting only minimal growth of cooling instabilities for this wire material. In contrast, the cooling time for tungsten is short (around 8 ns) compared to the other timescales, so the radiative cooling instability can grow significantly.

Due to the magnetic field advected by the plasma flows, which piles up at the conducting obstacle, we calculate that the electrons will be partially magnetised, and so thermal conduction will be reduced perpendicular to the magnetic field. This leads to a longer $\lambda_{Field}$ parallel to the magnetic field than perpendicular to it, which is supported by the shadowgraphy effects seen in the end-on interferometry images. Due to a lack of high-resolution, validated cooling curves for high-Z elements such as tungsten, it is not possible to precisely simulate this experiment or to calculate the predicted instability thresholds or growth rates. As such, we cannot definitively say that we observe radiative cooling instability in the experiments with tungsten, but the data we have is consistent with this hypothesis.
The other two mechanisms we considered (ETI and kinetic effects) cannot be responsible for the density perturbations we observe at 350 ns after current start, but either could potentially seed perturbations at early times which are then grown by the radiative cooling instability.

The spatially, temporally, and angularly resolved measurements made using multiple, simultaneous diagnostics in a reproducible experiment make these data ideal for comparison to simulations and theory, especially if suitably validated cooling curves are available. It also represents the first real test of the new imaging refractometer, which has been used to estimate the length scale of the perturbations. In future, we hope to develop a more complete theoretical model of the imaging refractometer that would allow us to relate higher-order moments of the distribution of deflection angles (such as the kurtosis) to the spectrum of density fluctuations within the plasma.

\section*{Acknowledgements}
We would like to thank Dr Alexander Rososhek and Dr Archie F. A. Bott for useful discussions regarding this data and diagnostic interpretation.
This work was supported by the Engineering and Physical Sciences Research Council (EPSRC) Grant No. EP/N013379/1 and the US Department of Energy (DOE), including Awards No. DE-NA0003764 and No. DE-SC0020434.
\bibliography{FinalAIPElectronDensityPerturbationsRefractometerAIPSubFile} 
\end{document}